\documentclass[a4paper, 11pt]{article}
\usepackage{amssymb,amsmath,graphicx}
\usepackage{amsfonts,textcomp}
\usepackage{natbib}
\usepackage[caption=false]{subfig}
\usepackage{epsfig,latexsym,graphicx}
\newtheorem{theorem}{Theorem}[section]

\newtheorem{proposition}[theorem]{Proposition}
\newtheorem{corollary}[theorem]{Corollary}
\newtheorem{remark}{Remark}[section]

\begin{document}

\title{
A New Family of Divergences Originating from Model Adequacy Tests 
and Application to Robust Statistical Inference
}

\author{Abhik Ghosh and Ayanendranath Basu\\
Interdisciplinary Statistical Research unit\\
Indian Statistical Institute \\
203 B. T. Road, Kolkata 700 108, India\\
{\it abhianik@gmail.com, ayanbasu@isical.ac.in}}
\date{}
\maketitle

\begin{abstract}

Minimum divergence methods are popular tools in a variety of statistical applications. 
We consider tubular model adequacy tests, and demonstrate
that the new divergences that are generated in the process are very 
useful in robust statistical inference. In particular we show that 
the family of $S$-divergences can be alternatively developed
using the tubular model adequacy tests; a further application of the 
paradigm generates a larger superfamily of  
divergences. We describe the properties of this 
larger class and its potential applications in robust
inference. Along the way, the failure of the first order influence function analysis 
in capturing the robustness of these procedures is also established. 

\end{abstract}

\section{Introduction and Preliminaries}\label{SEC:intro}

Divergence measures are frequently used in statistics and information theory. 
Minimum divergence techniques form an important component of modern statistical analysis
because of their strong robustness properties against outlying observations. 
Many density-based statistical divergences produce highly robust estimators along with  
high (sometimes even full) asymptotic efficiency and hence constitute a large class of useful generalizations of 
the extremely sensitive classical maximum likelihood estimator (MLE).
Popular examples of such divergences include the Kullback-Leibler divergence (KLD),
the Bregman divergence \citep{Bregman:1967}, 
the power divergence (PD) family of \cite{Cressie/Read:1984},
the density power divergence (DPD) family of \cite{Basu/etc:1998},
the generalized Kullback-Leibler (GKL) family of \cite{Park/Basu:2003},
the $S$-divergence (SD) family of \cite{Ghosh/etc:2016} and many more;
see  \cite{Beran:1977}, \cite{Simpson:1987}, 
\cite{Lindsay:1994}, \cite{Pardo:2006}, \cite{Basu/etc:2011} for details.
Some of these works have provided generalizations of existing families of divergences 
which extended the scope of statistical inference at that point of time
providing better trade-off between the dual goals in parametric estimation -- 
efficiency under the model and robustness away from it.

However, search for such divergences has often been ad-hoc, 
and researchers have largely relied on intuitive generalizations or empirical evidence.  
In the present paper, we study a systematic procedure of generating new (larger families of) divergences 
through suitable model adequacy tests (MATs) based on existing divergences. 
The MATs were originally proposed as alternatives to the chi-square type 
goodness-of-fit tests, since the latter group generally fails to pick out any specific model for sufficiently large 
sample sizes \citep{Liu/Lindsay:2009}. 
It may be that the specified model (simple and easy to work with)
is already very close to the true data generating distribution, 
yet these chi-square type goodness-of-fit tests might reject the simpler model at large sample sizes 
in favor of more complicated models which have marginally lower discrepancy but are otherwise hard to work with. 
But, the MATs allow us to test if the simpler specified model is within practically acceptable range of the true density 
\citep{Rudas/etc:1994,Hodges/Lehmann:1954,Goutis/Robert:1998,Dette/Munk:2003a,Liu/Lindsay:2009};
see also \cite{Rieder:1978,Rieder:1980}, \cite{Bickel:1984} and \cite{Donoho/Johnstone:1989}.
Here, for our purpose, we restrict ourselves to the MATs based on divergence measures only.

As a general terminology, let us use the term ``density" for both discrete and continuous probability 
mass or density function respectively and consider the class $\mathcal{G}$ of all absolutely continuous 
densities  with respect to some common dominating measure $\mu$. 
Define a statistical divergence $\rho(g,f)$ between two densities $g$ and $f$ 
as a non-negative function from $\mathcal{G}\times \mathcal{G}$ to $[0, \infty)$ 
which equals zero if and only if $g=f$, identically.
For such a divergence, we say that the assumed model family $\mathcal{F} =\{f\}$ 
is adequate at level $c$ with respect to the true probability density $g$ 
if the divergence (or, loosely, distance) of $g$ from $\mathcal{F}$, defined as
$
\rho(g, \mathcal{F}) = \displaystyle\inf_{f\in\mathcal{F}}\rho(g, f),
$
is less than or equal to $c$. Then the general MAT considers the null hypothesis
\begin{eqnarray}
H_0~:~\rho(g, \mathcal{F}) \leq c.
\label{EQ:MAT}
\end{eqnarray}
Given $c$, one can construct suitable tests for the problem of model adequacy; 
however, we further restrict to the divergence based tests including the likelihood ratio test (LRT). 
\cite{Liu/Lindsay:2009} constructed the tolerance region in (\ref{EQ:MAT}) by taking $\rho$ as the KLD measure given by
\begin{equation}
{\rm KLD}(g, f) = \int f \log(f/g)d\mu.
\label{EQ:KLD}
\end{equation}
Note that the KLD is adjoint to the popular likelihood divergence (LD) given by 
\begin{equation}
{\rm LD}(g, f) = \int g \log(g/f)d\mu,
\label{EQ:LD}
\end{equation}
obtained by reversing the roles of $f$ and $g$. 
\cite{Liu/Lindsay:2009} considered the LRT statistic (based on the LD) with $\rho = {\rm KLD}$ 
to illustrate several nice properties of this approach. 
Specifically and interestingly, their approach independently yields the GKL family 
that had already been shown to produce highly robust estimators by \cite{Park/Basu:2003}. 
This one parameter GKL family 
can be obtained  as the solution to the optimization problem (\ref{EQ:GKL_gen}) below
which arises in the construction of the LRT  for the MAT of \cite{Liu/Lindsay:2009}.
\begin{eqnarray}
{\rm GKL}_\tau(g,f) &=& \int \left[\frac{g}{\bar{\tau}}\log\left(\frac{g}{f}\right) 
- \left(\frac{g}{\bar{\tau}} + \frac{f}{\tau}\right)\log\left(\tau\frac{g}{f} + \bar{\tau}\right)\right]d\mu, 
~~~\bar{\tau} = 1-\tau,~\tau \in (0,1),\\
&=& \frac{1}{\tau\bar{\tau}} \min_{p\in\mathcal{G}}\left[\tau {\rm LD}(g,p)
+ \bar{\tau}{\rm KLD}(p,f)\right], 
\label{EQ:GKL_gen}
\label{EQ:GKL}
\end{eqnarray}
This nice interpretation of the GKL family has also been briefly noted by \cite{Park/Basu:2003} independently.
Further, they have indicated another MAT, obtained by reversing the role of KLD and the LD, 
which generates the Cressie-Read PD family given by 
\begin{eqnarray}
{\rm PD}_\lambda( g, f) &=& \frac{1}{\lambda(\lambda + 1)} \int  g [ ( g/f)^\lambda - 1 ]d\mu =
\label{EQ:PD}
\frac{1}{\tau\bar{\tau}}\min_{p\in\mathcal{G}}\left[\tau {\rm KLD}(g,p)
+ \bar{\tau}{\rm LD}(p,f_0)\right], ~~~
\label{EQ:MAT_PD}
\end{eqnarray}
with $\bar{\tau}=1-\tau = -\lambda$.
This PD family also has several useful applications in robust statistical inference \cite[see, e.g.,][]{Basu/Lindsay:1994, Basu/etc:2011}.

The above discussion invokes a natural question:  can we obtain larger superfamilies of statistical divergences 
that could be further helpful in robust inference like the PD or the GKL families, 
through more general MATs? In this paper, we present an answer by exploring some MATs based on more general divergences. 
In Section \ref{SEC:Inerpret_SD} we consider the recent two-parameter $S$-divergence (SD) family 
of \cite{Ghosh/etc:2016}, and illustrate its development through a suitable MAT.
Then,  we further generalize our MAT by replacing KLD and LD with appropriate members of the SD family
and explore its properties in Section \ref{SEC:MAT_SD}.
As expected, this general SD based MATs generate another {\it new} and 
larger super-family of divergences which, interestingly, contains both the SD and the GKL family as its subclasses.
This is another novel discovery within the literature of the density based divergences and the related robust statistical inference.
We refer to this larger three parameter superfamily of divergences as the Generalized $S$-divergence (GSD) family
and explore some of its basic properties in Section \ref{SEC:GSD}.
The potential applications of this GSD superfamily in robust estimation have been investigated in Section \ref{SEC:GSD_Inf}.
Section \ref{SEC:Illustratios} presents some numerical illustrations for the usefulness of the GSD based inference
and the paper ends with some concluding remarks in Section \ref{SEC:Conclusion}.

Before concluding this section, we summarize the major contributions of this paper as follows.
\begin{itemize}
\item We explore a systematic process of generating new divergence families from suitable model adequacy tests
and study the relation between them.

\item We demonstrate that the $S$-divergence family
can be generated through an appropriate model adequacy tests in the same spirit. 

\item 
A further application of this technique leads to the generation of an entirely new
family of divergences (GSD) containing the SD and the GKL families as special cases.

\item We acknowledge that there already exists too many divergences that have found little 
or no applications. However, an exploration of the properties of our newly developed superclass of GSD demonstrates that 
the divergences within this superclass, which provide the best combinations of robustness and efficiency,
do not belong to any of its previously studied subclasses. 
This indicates that studying the properties of the individual subclasses may not be sufficient 
to identify the ``best" divergence in respect of robust parametric estimation, 
so that this super-class does make some positive value addition to the literature. 

\item A study of the minimum divergence estimators generated by the GSD superclass reveals the
limitation of the first order influence function analysis in assessing their robustness.
\end{itemize}

\section{$S$-Divergence Family from A Suitable Model Adequacy Test}\label{SEC:Inerpret_SD}

The $S$-divergence family of density-based divergences, recently developed by \cite{Ghosh/etc:2016}, 
is a two-parameter $(\alpha, \lambda)$ generalization of the PD family that connects each member of the PD family 
(having parameter $\lambda$) at $\alpha = 0$ to the $L_2$ divergence at $\alpha=1$. This family is defined as 
\begin{equation}
S_{(\alpha, \lambda)}(g,f) =  \int \left[\frac{1}{A} f^{1+\alpha} - \frac{1+\alpha}{AB}f^{B} g^{A} + \frac{1}{B}g^{1+\alpha}\right]d\mu, 
\label{EQ:S_div_gen}
\end{equation}
where $A = 1+\lambda (1-\alpha)$ and  $B = \alpha - \lambda (1-\alpha)$. Clearly, $ A+B=1+\alpha $. 
Also the above form is defined only when $A \ne 0$ and $B \ne 0$;
for $A=0$ or $B=0$ the corresponding SD measure is defined by the continuous 
limit of (\ref{EQ:S_div_gen}) as $A \rightarrow 0$ or $B\rightarrow 0$ respectively.
Further, at the choice $\lambda = 0$, this SD family contains another popular divergence family, 
namely the density power divergence (DPD) family of \cite{Basu/etc:1998} having parameter $\alpha> 0$,
defined as
\begin{equation}
{\rm DPD}_\alpha(g,f) =  \int f^{1+\alpha}d\mu - \frac{1+\alpha}{\alpha}\int f^{\alpha} gd\mu + \frac{1}{\alpha}\int g^{1+\alpha}d\mu.
\label{EQ:DPD}
\end{equation}
In general, the SD measures are not symmetric. But it becomes symmetric if and only if 
either $\alpha = 1$ (which generates the $L_2$ divergence) or $\lambda = -\frac{1}{2}$. 
The latter case represents an interesting subclass,
referred to as the $S$-Hellinger Distances (SHD) in \cite{Ghosh/etc:2016}, given by
\begin{equation}
{\rm SHD}_\alpha(g,f) = S_{(\alpha, \lambda = -1/2)}(g, f)  =   \frac{2}{1+\alpha} \int 
\left( g^{(1+\alpha)/2} - f^{(1+\alpha)/2}\right)^2.
\label{EQ:SHD}
\end{equation}
It connects the Hellinger distance at $\alpha=0$ to $L_2$ divergence at $\alpha=1$.
Note that just as the Hellinger distance represents the self 
adjoint member of the PD family, any other cross section of the class of $S$-divergences 
for a fixed value of $\alpha$ has a self adjoint member in $S_{(\alpha, -1/2)}$.

The applications of SD family in robust parametric inferences have been described in 
\cite{Ghosh:2015}, \cite{Ghosh/Basu:2015, Ghosh/Basu:2016} and \cite{Ghosh/etc:2015, Ghosh/etc:2016}. 
It has been illustrated that some members of the SD family generate more robust inference
compared to its existing members within the PD and DPD subfamilies.
However, in terms of \cite{Broniatowski/TV:2012}, the SD measures are not decomposable, except for 
$\lambda=0$ (the DPD) or $\alpha=1$ (the $L_2$ divergence),
and hence require the use of a non-parametric density estimator for performing minimum divergence estimation 
under continuous models. A possible alternative approach may be developed along the line of \cite{Broniatowski/Keziou:2009}
and \cite{Toma/Broniatowski:2010}.

Although the SD family is seen to perform very well in robust statistical inference,
as one referee pointed-out, its development was mostly ad-hoc. 
In this paper, we demonstrate that one can obtain the SD family from a suitable MAT 
extending the arguments of \cite{Liu/Lindsay:2009},
which justify the systematic development of this useful divergence family with mathematical rigor. 
Since the SD family is an extension of the PD family,
we intuitively should be able to obtain it as solution of an appropriate generalization of the problem in (\ref{EQ:MAT_PD})
using extended KLD and LD measures through the parameter $\alpha$.
In particular, we consider the extended families given by
\begin{eqnarray}
{\rm SKL}_\alpha(g,f) &=& \int f^{1+\alpha} \log\left(\frac{f}{g}\right)d\mu - \int \frac{(f^{1+\alpha} - g^{1+\alpha})}{{1+\alpha}}d\mu,
~~~~~\alpha \geq 0,
\label{EQ:SKL_defn}\\
{\rm SLD}_\alpha(g,f) &=& \int g^{1+\alpha} \log\left(\frac{g}{f}\right)d\mu - \int \frac{(g^{1+\alpha} - f^{1+\alpha})}{{1+\alpha}}d\mu,
~~~~~\alpha \geq 0.
\label{EQ:SLD_defn}
\end{eqnarray}
Note that, at $\alpha=0$, the above families simplify, respectively, to the KLD and the LD measures.
They are also the particular subfamilies within the SD family corresponding to $A=0$ and $B=0$ respectively.
Further, \cite{Liu/Lindsay:2009} have noted that the curvature of the KLD and the LD cancel each other 
to generate the new family of divergence measures; this is because they are adjoint to each other
in the language of \cite{Jimenez/Shao:2001}. 
This special geometry also holds for our SLD and SKL families at any given $\alpha$, i.e., 
$$
{\rm SKL}_\alpha(g,f) = {\rm SLD}_\alpha(f,g), ~~f,g\in \mathcal{G}, ~\alpha\in [0,1).
$$
Further, they are also symmetrically opposite to the SHD family, the only self-adjoint subfamily within the SD family.
These properties of the SKL and SLD families intuitively indicate that the use of the members of the SKL and SLD families 
at any particular $\alpha\in [0,1)$  in constructing the MAT would generate the 
corresponding cross-section of the SD family with the same $\alpha$.


To prove this intuition rigorously, let us fix an $\alpha\in [0,1)$ and a tolerance limit $c$. 
We want to test whether our model is adequate 
in terms of the SLD$_\alpha$ measure at level $c$. This is equivalent to test for the null hypothesis
\begin{eqnarray}
H_0: {\rm SLD}_\alpha(g,\mathcal{F})\leq c.
\label{EQ:2null}
\end{eqnarray}
Let us construct a divergence based test (rather than the LRT)
for this null hypothesis  using the SKL measure with the  same tuning parameter $\alpha$. 
These divergence based tests are quite robust compared to the LRT as explored in \cite{Ghosh/etc:2015} and \cite{Ghosh/Basu:2016}.
Then, based on a sample of size $n$,  the resulting test statistic would be $2n {\rm SKL}_\alpha(g,\hat{f}_c)$,
where $\hat{f}_c$ is the estimated density under the null hypothesis (\ref{EQ:2null}) 
obtained by minimizing the SKL measure between the true density $g$ and the model $f$ 
over all possible model densities satisfying (\ref{EQ:2null}). 
Thus, $\hat{f}_c$ is nothing but the solution of the constrained optimization problem
\begin{equation}
\min_{p\in\mathcal{G}}{\rm SKL}_\alpha(g,p) ~~~\mbox{subject to }~~~{\rm SLD}_\alpha(p, \mathcal{F})\leq c.
\label{EQ:constrained1_SD}
\end{equation}

As in \cite{Liu/Lindsay:2009}, we transfer this constrained optimization problem to an equivalent simpler
unconstrained optimization problem. 
For simplicity, let us first assume that the model family contains only one element, i.e., $\mathcal{F}=\{f_0\}$. 
Then,  (\ref{EQ:constrained1_SD}) simplifies to 
\begin{equation}
\min_{p\in\mathcal{G}}{\rm SKL}_\alpha(g,p) ~~~\mbox{subject to }~~~{\rm SLD}_\alpha(p, f_0)\leq c.
\label{EQ:constrained2_SD}
\end{equation}

\begin{theorem}
The constrained optimization problem (\ref{EQ:constrained2_SD}) has a unique solution, on the boundary, 
having the form
\begin{equation}
\hat{f}_c=\hat{f}_\tau = g^\tau  f_0^{(1-\tau)},
\label{EQ:sol_form_SD}
\end{equation}
for some $\tau\in[0,1]$, a function of given $c$,
defined through the boundary condition 
\end{theorem}
\begin{equation}
{\rm SLD}_\alpha(\hat{f}_\tau, f_0)= c.
\label{EQ:BC}
\end{equation}
\label{THM:Solution_SD_MAT}

Further, starting from a fixed $\tau\in[0,1]$ we can see that the simpler unconstrained problem 
\begin{equation}
\min_{p} \left[\tau {\rm SKL}_\alpha(g,p) + \bar{\tau}{\rm SLD}_\alpha(p, f_0)\right], ~~\bar{\tau} = 1- \tau,
\label{EQ:Unconstrained_SD}
\end{equation}
has the solution $\hat{f}_\tau$ of the same form as (\ref{EQ:sol_form_SD}).
The formulations presented earlier in Section \ref{SEC:intro} through (\ref{EQ:GKL_gen}) 
and (\ref{EQ:MAT_PD}) are exactly in the same spirit for the particular cases when the divergences are the LD and the KLD. 
As argued in \cite{Liu/Lindsay:2009}, there is a direct one to one correspondence between the parameters $\tau$ and $c$
in the two optimization problems in (\ref{EQ:Unconstrained_SD}) and (\ref{EQ:constrained2_SD}) respectively
through (\ref{EQ:BC}).
Thus the targeted quantity $\hat{f}_c$ in our MAT statistic $2n {\rm SKL}_\alpha(g,\hat{f}_c)$
can be easily obtained as the solution to the simpler unconstrained optimization problem in (\ref{EQ:Unconstrained_SD})
for a suitable $\tau$-value satisfying (\ref{EQ:BC}).

Now, suppose the model family $\mathcal{F}$ contains more than one element indexed by the parameter $\theta$ as 
$\mathcal{F} = \{f_\theta\}$. Then, we should extend the unconstrained problem in (\ref{EQ:Unconstrained_SD}) to
the problem 
\begin{equation}
\min_{p} \left[\tau {\rm SKL}_\alpha(g,p) + \bar{\tau} {\rm SLD}_\alpha(p, \mathcal{F})\right]
\equiv \min_\theta\min_{p} \left[\tau {\rm SKL}_\alpha(g,p) + \bar{\tau} {\rm SLD}_\alpha(p, f_\theta)\right].
\label{EQ:Unconstrained2_SD}
\end{equation}
Again we can show that the required solution to the constrained optimization problem in (\ref{EQ:constrained1_SD})
is nothing but the solution to this unconstrained problem in (\ref{EQ:Unconstrained2_SD}) with some appropriate value of $\tau$
satisfying  $c={\rm SLD}_\alpha(\hat{f}_\tau, \mathcal{F})=  \min_\theta {\rm SLD}_\alpha(\hat{f}_\tau, f_\theta)$.
Further, the inner minimization in (\ref{EQ:Unconstrained2_SD}) is of the form (\ref{EQ:Unconstrained_SD}) for a given $\theta$
and has the explicit solution $\hat{f}_\tau = g^{\tau} f_\theta^{\bar{\tau}}$.
Substituting it, (\ref{EQ:Unconstrained2_SD}) becomes
$\displaystyle\min_\theta \left[\tau {\rm SKL}_\alpha(g,\hat{f}_\tau) + \bar{\tau} {\rm SLD}_\alpha(\hat{f}_\tau, f_\theta)\right].$
However, some routine algebra shows that
\begin{eqnarray}
 \left[\tau {\rm SKL}_\alpha(g,\hat{f}_\tau) + \bar{\tau} {\rm SLD}_\alpha(\hat{f}_\tau, f_\theta)\right]
= \left[\tau {\rm SKL}_\alpha(g,g^\tau f_\theta^{\bar{\tau}}) + \bar{\tau} {\rm SLD}_\alpha(g^\tau f_\theta^{\bar{\tau}}, f_\theta)\right]
=\tau \bar{\tau}S_{(\alpha, \lambda_\tau)}(g,f_\theta),\nonumber
\end{eqnarray}
where $\bar{\tau} = 1 -\tau$ and $\lambda_\tau$ is a function of $\tau$ defined as
\begin{eqnarray}
\lambda_\tau = \frac{\alpha\tau-\bar{\tau}}{1-\alpha}.
\label{EQ:lambda_tau}
\end{eqnarray}
Therefore, our required solution $\hat{f}_\tau$ is indeed the model element $f_{\hat{\theta}_\tau}$
with $\hat{\theta}_\tau$  being the minimum SD estimator (MSDE) defined as
$
\hat{\theta}_\tau = \displaystyle\arg\min_\theta S_{(\alpha, \lambda_\tau)}(g,f_\theta).
$
Thus, our MAT based on SKL$_\alpha$ independently yields the form of the SD measure with appropriate parameters
and directly depends on the corresponding MSDE 
having nice robustness properties \citep{Ghosh/etc:2016}.

Further, noting that, given $\alpha<1$, $\lambda_\tau$ is a one-to-one function of $\tau$, 
and any SD measure with parameter $\alpha\in[0,1)$ and $\lambda \in [0, \frac{\alpha}{1-\alpha}]$ 
can be obtained from the MAT based on the SKL with the same $\alpha$ and
an appropriate value of $c$ obtained from $\tau = \frac{\lambda(1-\alpha) + 1}{\alpha+1}$.
Therefore the SD family has a direct one to one correspondence with the MAT based on SKL and SLD.


\section{$S$-Divergence based General Model Adequacy Tests}\label{SEC:MAT_SD}

The nice and interesting interpretation of the $S$-divergence measures as discussed in the previous section invokes the next question:
what if we start with a more general divergence family instead of the SLD and SKL? 
Will we get an even  larger superfamily of divergences in that case?
To answer these questions, we now develop a general MAT based on the SD family.

Noting the special geometric curvatures of the SKL and SLD families and the fact that the SD family is
symmetric about $\lambda = - \frac{1}{2}$, we may try to replace the SKL and SLD by 
suitable members of the general SD family. 
More precisely, for any $\alpha\in[0,1)$, we want to consider two SD measures with parameters $(\alpha,\lambda')$
and $(\alpha,\lambda'')$ in place of SLD$_\alpha$ and SKL$_\alpha$ in such a way that their geometric curvatures 
cancel each other; i.e.,  these two members should be symmetrically opposite with respect to  the line $\lambda=-1/2$ and satisfy
$$
S_{(\alpha, \lambda')}(g,f) = S_{(\alpha, \lambda'')}(f, g), ~~~~ \forall ~ f, g \in \mathcal{G}.
$$
From the derivations of the previous section, we may take $\lambda'=\frac{\gamma}{1-\alpha}$ for some $\gamma\in \mathbb{R}$
and then the above requirements give us the choice $\lambda'' =  \frac{\alpha-1-\gamma}{1-\alpha}$.
So, we consider the null hypothesis
\begin{equation}
H_0: S_{(\alpha, \frac{\alpha-1-\gamma}{1-\alpha})}(g, \mathcal{F}) \leq c,
\label{EQ:MAT_gen}
\end{equation}
and construct the MAT based on $S_{(\alpha, \frac{\gamma}{1-\alpha})}(f, g)$.
Note that the choice $\gamma=-1$ yields the MAT of Section \ref{SEC:Inerpret_SD} based on the SKL and the SLD measures. 
Based on a sample of size $n$, we define the MAT statistics for the hypothesis (\ref{EQ:MAT_gen}) as given by 
$T_n = 2n S_{(\alpha, \frac{\gamma}{1-\alpha})}(g, \hat{f}_c),$ 
where $\hat{f}_c$ is the best model element (with respect to the $S_{(\alpha, \frac{\gamma}{1-\alpha})}$ measure)
under the restriction imposed by $H_0$ in (\ref{EQ:MAT_gen}).
Thus, the $\hat{f}_c$ can be obtained as the solution of the constrained optimization problem
\begin{equation}
\min_{p}S_{(\alpha, \frac{\gamma}{1-\alpha})}(g, p) ~~~\mbox{subject to }~~~S_{(\alpha, \frac{\alpha-1-\gamma}{1-\alpha})}(p, \mathcal{F}) 
\leq c.
\label{EQ:constrained1_gen}
\end{equation}

Proceeding as in the previous section, we can again show that the unique solution to the above constrained optimization
problem with general model family $\mathcal{F}=\{f_\theta\}$ is nothing but
the unique solution $\hat{f}_\tau$ of the unconstrained optimization problem 
\begin{equation}
\min_{p\in\mathcal{F}}\left[\tau S_{(\alpha, \frac{\gamma}{1-\alpha})}(g,p)
+ \bar{\tau}S_{(\alpha, \frac{\alpha-1-\gamma}{1-\alpha})}(p,\mathcal{F})\right]
\equiv \min_\theta\min_{p}\left[\tau S_{(\alpha, \frac{\gamma}{1-\alpha})}(g,p)
+ \bar{\tau}S_{(\alpha, \frac{\alpha-1-\gamma}{1-\alpha})}(p,f_\theta)\right],
\label{EQ:Unconstrained_gen}
\end{equation}
for some $\tau\in [0,1]$ that depends on $c$ through 
$
c= S_{(\alpha, \frac{\alpha-1-\gamma}{1-\alpha})}(\hat{f}_\tau,\mathcal{F})$ 
$= \displaystyle\min_\theta S_{(\alpha, \frac{\alpha-1-\gamma}{1-\alpha})}(\hat{f}_\tau,f_\theta).
$
To see this, we again consider the singleton model family $\mathcal{F}=\{f_0\}$. 
Then, the following theorem presents the solution of the simplified version of the constrained problem (\ref{EQ:constrained1_gen}) 
given by
\begin{equation}
\min_{p}S_{(\alpha, \frac{\gamma}{1-\alpha})}(g, p) ~~~\mbox{subject to }~~~S_{(\alpha, \frac{\alpha-1-\gamma}{1-\alpha})}(p, f_0) 
\leq c.\label{EQ:constrained2_gen}
\end{equation}
The proof of the theorem follows from the method of Lagrange multiplier and is hence omitted.

\begin{theorem}
	The constrained optimization problem (\ref{EQ:constrained2_gen}) has a unique solution, on the boundary,
	having the form
\begin{eqnarray}
\hat{f}_c =\hat{f}_\tau &=& \left[\tau g^{1+\gamma} + (1-\tau) f_0^{1+\gamma}\right]^{\frac{1}{1+\gamma}}, 
~~~\mbox{if  }~\gamma\neq -1,	\label{EQ:sol_form_gen}\\
 &=& g^\tau f_0^{(1-\tau)}, ~~~~~~~~~~~~~~~~~~~~~~~~~ \mbox{if  }~\gamma=-1,
\end{eqnarray}
	for some $\tau\in[0,1]$, a function of given $c$, defined through the condition
	$S_{(\alpha, \frac{\alpha-1-\gamma}{1-\alpha})}(\hat{f}_\tau,f_0) = c$.
	\label{THM:Solution_gen_MAT}
\end{theorem}

However, we can immediately check that the solution $\hat{f}_\tau$ as given in the above theorem is also 
the unique solution to the much simpler unconstrained optimization problem 
\begin{equation}
\min_{p\in\mathcal{F}}\left[\tau S_{(\alpha, \frac{\gamma}{1-\alpha})}(g,p)
+ \bar{\tau}S_{(\alpha, \frac{\alpha-1-\gamma}{1-\alpha})}(p,f_0)\right],
\label{EQ:Unconstrained1_gen}
\end{equation}
for the $\tau$ value satisfying $S_{(\alpha, \frac{\alpha-1-\gamma}{1-\alpha})}(\hat{f}_\tau,f_0) = c$.
Now, for the general model family $\mathcal{F}=\{f_\theta\}$, we need to extend this to the
problem in (\ref{EQ:Unconstrained_gen}) which again has the same solution as to the constrained optimization problem 
in (\ref{EQ:constrained1_gen}). Clearly, the inner minimization in (\ref{EQ:Unconstrained_gen}) 
has an explicit solution $\hat{f}_\tau$ as given in Theorem \ref{THM:Solution_gen_MAT} 
and hence the optimization problem in (\ref{EQ:Unconstrained_gen}) reduces to the form
\begin{equation}
\min_\theta \left[\tau  S_{(\alpha, \frac{\gamma}{1-\alpha})}(g,\hat{f}_\tau) 
+ \bar{\tau} S_{(\alpha, \frac{\alpha-1-\gamma}{1-\alpha})}(\hat{f}_\tau, f_\theta)\right], 
~~~~\bar{\tau} = 1 -\tau.
\label{EQ:Unconstrained3_gen}
\end{equation}
For $\gamma\neq -1$, we can simplify the argument in the above optimization problem in (\ref{EQ:Unconstrained3_gen}) as
$\left[\tau  S_{(\alpha, \frac{\gamma}{1-\alpha})}(g,\hat{f}_\tau) 
+ \bar{\tau} S_{(\alpha, \frac{\alpha-1-\gamma}{1-\alpha})}(\hat{f}_\tau, f_\theta)\right]
= \tau \bar{\tau}Q_{(\alpha,\gamma,\tau)}(g,f_\theta),$
where $Q_{(\alpha,\gamma,\tau)}(g,f)$ is defined as
\begin{equation}
Q_{(\alpha,\gamma,\tau)}(g,f) = \frac{1}{\tau\bar{\tau}(\alpha-\gamma)}
\int \left[\left\{\tau\left(\frac{g}{f}\right)^{1+\alpha} +\bar{\tau}\right\} - 
\left\{\tau\left(\frac{g}{f}\right)^{1+\gamma} +\bar{\tau}\right\}^{\frac{1+\alpha}{1+\gamma}}\right]f^{1+\alpha}d\mu.
\label{EQ:Def0_GSD}
\end{equation}
The case $\gamma=-1$ is identical to that considered in Section \ref{SEC:Inerpret_SD} and hence leads to the appropriate
member of the SD family which can also be shown to be equal to 
$\displaystyle\lim_{\gamma\rightarrow -1} Q_{(\alpha,\gamma,\tau)}(g,f)$.
Then the required solution $\hat{f}_\tau$ is given by the model element $f_{\hat{\theta}_\tau}$ with 
$
\hat{\theta}_\tau = \arg\min_\theta Q_{(\alpha, \gamma, \tau)}(g,f_\theta).
$

Thus, our general MAT based on the SD measures yields
a new divergence function (see Theorem \ref{THM:GSD_pro1}) through the quantity $Q_{(\alpha,\gamma,\tau)}(g,f)$. 
This function $Q_{(\alpha,\gamma,\tau)}(g,f)$ can be easily extended to the cases  $\tau=0,1$ and $\alpha=\gamma$
through continuous limits of (\ref{EQ:Def0_GSD}).
Further the choice $\gamma=-1$ reduces the general divergence measure $Q_{(\alpha,\gamma,\tau)}(g,f)$ to the 
suitable SD measure with parameters $\alpha$ and $\lambda_\tau$ (defined in (\ref{EQ:lambda_tau})). 
So, we refer to this general three parameter family as the ``Generalized Super-Divergence (GSD) Family".

Note that given the generated sample data, we can easily compute the MAT statistics $T_n$ by using the 
minimum GSD estimators (MGSDEs) of the model parameter $\theta$. 
Hence the performance and robustness of this general MAT directly depend  on that of the MGSDEs.
In the rest of this paper, we illustrate the properties of the GSD and the MGSDEs in detail
and leave the exploration of the general MAT 
for our future research work.


\section{The Generalized $S$-Divergence Family and Basic Properties}\label{SEC:GSD}

The generalized $S$-divergence (GSD) family, as derived in the previous section, is a three parameter family 
of statistical divergences $Q_{(\alpha,\gamma,\tau)}(g,f)$ defined in (\ref{EQ:Def0_GSD}) 
for $\alpha\in[0,1)$, $\tau\in(0,1)$ and $\gamma\in\mathbb{R}-\{-1\}$ with $\gamma\neq \alpha$ and 
can be extended over $\alpha\in[0,1]$, $\tau\in[0,1]$ and $\gamma\in\mathbb{R}$ through their continuous limits in (\ref{EQ:Def0_GSD}).
In Table \ref{TAB:spcl}, we have listed several divergences or families of divergences which belong to the GSD family; 
they are mostly obtained as limiting members.

 \begin{table}[h]
 	\centering 
 	\caption{Existing Divergences as Special Cases of the GSD Family}
 	\begin{tabular}{l|l} \hline
 Tuning Parameters & Divergences\\
 (including limiting values) & (with reduced parameters if appropriate)\\\hline\hline
 $\alpha$, $\gamma\rightarrow -1$, $\tau$ & $S$-Divergence  (Eq.\ref{EQ:S_div_gen}) with $\alpha$, $\lambda_\tau$ in (Eq.\ref{EQ:lambda_tau})\\
 $\alpha$, $\gamma$, $\tau\rightarrow 0$ & $S$-Divergence  (Eq.\ref{EQ:S_div_gen}) with $\alpha$, $\frac{\gamma}{1-\alpha}$\\
 $\alpha$, $\gamma$, $\tau\rightarrow 1$ & $S$-Divergence  (Eq.\ref{EQ:S_div_gen}) with $\alpha$, $\frac{\alpha-1-\gamma}{1-\alpha}$\\
 \hline
 $\alpha$, $\gamma\rightarrow -1$, $\tau\rightarrow 0$ & SKL-Divergence  (Eq.\ref{EQ:SKL_defn}) with $\alpha$\\
 $\alpha$, $\gamma\rightarrow \alpha$, $\tau\rightarrow 1$ & SKL-Divergence  (Eq.\ref{EQ:SKL_defn}) with $\alpha$\\\hline
 $\alpha$, $\gamma\rightarrow -1$, $\tau\rightarrow 1$ & SLD  (Eq.\ref{EQ:SLD_defn}) with $\alpha$\\
 $\alpha$, $\gamma\rightarrow \alpha$, $\tau\rightarrow 0$ & SLD  (Eq.\ref{EQ:SLD_defn}) with $\alpha$\\\hline
 $\alpha=0$, $\gamma\rightarrow 0$, $\tau$ & GKL-Divergence  (Eq.\ref{EQ:GKL}) with $\tau$\\\hline
$\alpha=0$, $\gamma\rightarrow -1$, $\tau$ & PD (Eq.\ref{EQ:PD}) with $\lambda_\tau$ in (Eq.\ref{EQ:lambda_tau})\\\hline
$\alpha$, $\gamma\rightarrow -1$, $\tau=\frac{1}{1+\alpha}$ & DPD (Eq.\ref{EQ:DPD}) with $\alpha$\\\hline
 $\alpha=0$, $\gamma\rightarrow 0$, $\tau\rightarrow 0$ & LD (Eq.\ref{EQ:LD})\\
 $\alpha=0$, $\gamma\rightarrow -1$, $\tau\rightarrow 1$ & LD (Eq.\ref{EQ:LD})\\\hline
 $\alpha=0$, $\gamma\rightarrow 0$, $\tau\rightarrow 1$ & KLD (Eq.\ref{EQ:KLD})\\
 $\alpha=0$, $\gamma\rightarrow -1$, $\tau\rightarrow 0$ & KLD (Eq.\ref{EQ:KLD})\\\hline
 $\alpha$, $\gamma=\frac{\alpha-1}{2}$, $\tau$ & SHD (Eq.\ref{EQ:SHD}) with parameter $\alpha$\\\hline
 $\alpha=0$, $\gamma=-\frac{1}{2}$, $\tau$ & Hellinger Distance\\\hline
   		\hline 
 	\end{tabular}
 	\label{TAB:spcl}
 \end{table}

Thus, the GSD family contains the existing divergence families 
like the SD family (and hence the PD and DPD families) and the GKL family of \cite{Park/Basu:2003} 
as its limiting special cases and yields many more interesting new divergence measures.
One such interesting new subfamily arises at the limiting choice $\gamma\rightarrow\alpha$ 
which  has the form
\begin{eqnarray}
&& Q_{(\alpha,\alpha,\tau)}(g,f) = \lim\limits_{\gamma\rightarrow \alpha}  Q_{(\alpha,\gamma,\tau)}(g,f) \nonumber\\
&=& \int \frac{g^{1+\alpha}}{\bar{\tau}} \log\left(\frac{g}{f}\right) 
- \frac{1}{1+\alpha} \int \left(\frac{g^{1+\alpha}}{\bar{\tau}} + \frac{f^{1+\alpha}}{\tau}\right)
\log\left(\tau\left(\frac{g}{f}\right)^{1+\alpha} + \bar{\tau}\right)d\mu,
\end{eqnarray}
This subfamily is an interesting generalization of the GKL family in (\ref{EQ:GKL}), 
over the parameter $\alpha\in[0,1]$ and simplifies to the GKL family at $\alpha=0$.

First, we show that all the members of the GSD family represent proper statistical divergences. 
To see this, we can rewrite the GSD family in (\ref{EQ:Def0_GSD}) in terms of the Pearson residual $\delta =\frac{g}{f} - 1$ as
$Q_{(\alpha,\gamma,\tau)}(g,f) = \int C_{(\alpha,\gamma,\tau)}(\delta)f^{1+\alpha},$
where 
\begin{equation}
C_{(\alpha,\gamma,\tau)}(\delta) = \frac{1}{\tau\bar{\tau}(\alpha-\gamma)}\left[\left(\tau (\delta+1)^{1+\alpha} + \bar{\tau}\right)
-\left(\tau (\delta+1)^{1+\gamma} + \bar{\tau}\right)^{\frac{1+\alpha}{1+\gamma}}\right], ~~\delta\geq -1. 
\end{equation}
This function $C_{(\alpha,\gamma,\tau)}(\delta)$ is strictly convex  on $[-1, \infty)$ and satisfies the relations
$C_{(\alpha,\gamma,\tau)}(0) = 0 = C_{(\alpha,\gamma,\tau)}'(0)$ and $C_{(\alpha,\gamma,\tau)}''(0) = 1+\alpha>0$;
here $'$ and $''$ denote the first and second order derivatives respectively with respect to the argument.
Using them, we get the following theorem.

\begin{theorem}
All the functions $Q_{(\alpha,\gamma,\tau)}(g,f)$ in the generalized $S$-divergence family for $\alpha, \tau\in [0,1]$
and $\gamma\in\mathbb{R}$ define proper statistical divergences in the sense that, for any two densities $g,f\in \mathcal{G}$,
$Q_{(\alpha,\gamma,\tau)}(g,f) \geq 0$ with equality if and only if $g=f$.
\label{THM:GSD_pro1}
\end{theorem}

Examining the forms of the GSD measures, we get he following interesting properties.

\begin{theorem}
For any two densities $f,g\in\mathcal{G}$, the generalized 	$S$-divergence measure $Q_{(\alpha,\gamma,\tau)}(g,f)$
satisfies the following properties:
\begin{enumerate}
\item For any $\alpha,\gamma$, the GSD is adjoint with respect to $\tau=1/2$, i.e.,
$
Q_{(\alpha,\gamma,\tau)}(g,f) = Q_{(\alpha,\gamma,\bar{\tau})}(f,g).$ 
\item The GSD at $\tau=1/2$  is self-adjoint and symmetric in its arguments and given by
\begin{equation}
Q_{(\alpha,\gamma,1/2)}(g,f) = \frac{2}{(\alpha-\gamma)} \int\left[\left(g^{1+\alpha} + f^{1+\alpha}\right)
- 2^{-\frac{\alpha-\gamma}{1+\gamma}} \left(g^{1+\gamma} + f^{1+\gamma}\right)^{\frac{1+\alpha}{1+\gamma}}\right].
\end{equation}
\item The GSD measure also becomes symmetric and self-adjoint at $\gamma=\frac{\alpha-1}{2}$, 
where it becomes independent of the parameter $\tau$ and coincides with the SHD family.
\end{enumerate}
\label{THM:GSD_pro2}
\end{theorem}

From Table \ref{TAB:spcl} and Theorem \ref{THM:GSD_pro2}, we  observe that 
all members of the GSD family are not distinct; they (may) become identically equal at two or more 
parameter combinations $(\alpha,\gamma,\tau)$. For example, 
$Q_{(\alpha,-1,0)} = Q_{(\alpha,\alpha,1)}$, 
$Q_{(\alpha,-1,1)} = Q_{(\alpha,\alpha,0)}$ and  
$Q_{(\alpha,\frac{\alpha-1}{2},\tau)} = Q_{(\alpha,\frac{\alpha-1}{2},\tau')}$,
for all $\alpha,\tau, \tau' \in[0, 1]$.
Therefore, it will be an interesting future challenge to find the underlying 
geometry and topological properties of the  GSD family and 
hence characterize all its distinct members.

\section{Application of GSD Family in Robust Parametric Estimation}\label{SEC:GSD_Inf}

\subsection{The Set-up and Estimating Equation}\label{SEC:MGSDE_Set-up}

Let us consider the parametric model family of densities 
$\mathcal{F} = \{f_\theta : \theta \in \Theta \subset \mathbb{R}^p\}$. 
Let $F_\theta$ denote the distribution function corresponding to $f_\theta$.
We are interested in estimating parameter $\theta$. 
Let $G$ denote the distribution function corresponding to the true density $g$.
The MGSDE functional $T_{\alpha,\gamma,\tau}(G)$ at $G$ 
is then defined by the relation  
\begin{eqnarray}\label{EQ:MSDE_functional}
Q_{(\alpha,\gamma,\tau)}(g, f_{T_{\alpha,\gamma,\tau}(G)}) 
= \min_{\theta \in \Theta} Q_{(\alpha,\gamma,\tau)}(g, f_\theta),
\end{eqnarray}
provided the minimum exists. 
Some general conditions for existence of this MGSDE functional is given in Theorem \ref{THM:GSD_Func},
which follows from the definition of GSD and the properties of continuous functions over compact sets.
When $G$ is outside the model, $\theta_{\alpha,\gamma,\tau}^g = T_{\alpha,\gamma,\tau}(G)$ represents the best fitting parameter, and 
$f_{\theta^g}$ is the model element closest to $g$ in the GSD sense. 
For simplicity, we suppress the subscripts $\alpha, \gamma, \tau$ for $\theta_{\alpha,\gamma,\tau}^g$.

\begin{theorem}
Consider the MGSDE functional $T_{\alpha,\gamma,\tau}$ in (\ref{EQ:MSDE_functional}). Then, we have the following.
\begin{enumerate}
\item If $g=f_{\theta_0}\in \mathcal{F}$, then  the MGSDE functional $T_{\alpha,\gamma,\tau}(G)$ exists and equals $\theta_0$.
Further if the model family $\mathcal{F}$ is identifiable, $T_{\alpha,\gamma,\tau}$ is unique and Fisher consistent
on $\mathcal{F}$.

\item If $\Theta$ is compact and $f_\theta(x)$ is continuous in $\theta$ for almost all $x$, 
then the MGSDE functional $T_{\alpha,\gamma,\tau}(G)$ exists for all $g\in\mathcal{G}$.
\end{enumerate}\label{THM:GSD_Func}
\end{theorem}

Given the observed data, we estimate $\theta$ by minimizing 
the divergence $Q_{(\alpha, \gamma,\tau)}(\hat{g},f_{\theta})$ 
over $\theta \in \Theta$, where $\hat{g}$ is some non-parametric estimate of the true density $g$. 
When the model is discrete, a simple choice for $\hat{g}$ is given by the relative frequencies; 
for continuous models we need a nonparametric density estimator. 
Thus, the estimating equation for the MGSDE is given by 
\begin{eqnarray}
\int f_{\theta}^{1+\alpha} u_{\theta}d\mu 
- \int \left(\tau\hat{g}^{1+\gamma} + \bar{\tau}f_\theta^{1+\gamma}\right)^{\frac{\alpha-\gamma}{1+\gamma}}f_{\theta}^{1+\gamma} u_{\theta}d\mu &=& 0, \nonumber \\ 
\mbox{or} ~~~~ \int K(\delta(x))f_{\theta}^{1+\alpha}(x) u_{\theta}(x)d\mu(x) &=& 0, 
\label{EQ:S-divergence_est_equation}
\end{eqnarray}
where $\delta(x) = \delta_n(x) = \frac{\hat{g}(x)}{f_{\theta}(x)} - 1$ and 
$K(\delta) = \frac{1}{\tau(\alpha-\gamma)}
\left[\left(\tau(\delta+1)^{1+\gamma} + \bar{\tau}\right)^{\frac{\alpha-\gamma}{1+\gamma}} -1 \right]$.
Note that, the function $K(\delta)$ satisfies the relations $K(0) = 0$ and $K'(0)=1$.
Therefore, this estimating equation of the MGSDE is an {\it unbiased estimating equation} at the model $g=f_\theta$.
Further, at any fixed $\alpha$, the estimating equations of different MGSDEs differ only in terms of the function $K(\cdot)$, 
just like the case of MSDEs in \cite{Ghosh/etc:2016}.
Hence, the robustness properties of the resulting estimators are expected to depend at least partially 
on the form of this $K(\cdot)$ function. 
However, all divergences within the GSD family produce affine equivariant estimators.

\begin{proposition}
Consider the affine transformation $Y = UX + v$ for a non-singular matrix $U$ and 
a vector $v$ of the dimension as of $X$. Then we have 
	$
	Q_{(\alpha,\gamma,\tau)}(g_{_Y},f_{_Y}) = k Q_{(\alpha,\gamma,\tau)}(g_{_X}, f_{_X}), 
	$ 
	where $k = |Det(U)|^{1+\alpha} > 0$. 
	Therefore the divergence measure $Q_{(\alpha,\gamma,\tau)}(g,f)$ is not itself affine invariant,
	but the corresponding minimum divergence estimator is affine equivariant.
\end{proposition}

\subsection{Robustness properties: Influence Function}\label{SEC:S-DIV-IF}

The influence function (IF) is a popular and classical indicator of the first-order robustness and efficiency of any estimator. 
It indicates the asymptotic effect of the infinitesimal contamination on 
the properties of the estimator in the neighborhood of the true distribution $G$ \citep{Hampel:1974,Hampel/etc:1986}.
More precisely, the IF of any statistical functional $T(G)$ at $G$ is defined as 
$
IF(y, T, G) = \left.\frac{\partial}{\partial\epsilon}T(G_\epsilon)\right|_{\epsilon=0},
$
where $G_\epsilon = (1-\epsilon)G + \epsilon\wedge_y$ is the contaminated distribution,
$\epsilon$ is the contamination proportion and $\wedge_y$ denote the degenerate distribution at the contamination point $y$. 
The IF of a robust estimator should be bounded;
its non-boundedness indicates that the first order asymptotic bias of the estimator may diverge to infinity under contamination.

Now, let us consider the MGSDE functional $T_{\alpha,\gamma,\tau}(\cdot)$ 
as defined in (\ref{EQ:MSDE_functional}), which satisfies the estimating equation (\ref{EQ:S-divergence_est_equation})
with $\hat{g}$ replaced by $g$. Then, a straightforward differentiation of the estimating equation yields 
its IF as presented in the following theorem.

\begin{theorem}\label{THM:MSDE_IF}
Consider the general set-up of the MGSDE as in Section \ref{SEC:MGSDE_Set-up}.
Then, the influence function of the MGSDE functional $T_{\alpha,\gamma,\tau}$ is given by 
	\begin{equation}
	\label{Influence_Function_S_divergence}
	IF(y; T_{\alpha,\gamma,\tau}, G) = J_g^{-1} \left[ u_{\theta^g}(y) f_{\theta^g}^{1+\gamma}(y) g^{\gamma}(y)
	\left(\tau g^{1+\gamma}(y) + \bar{\tau}f_{\theta^g}^{1+\gamma}(y)\right)^{\frac{\alpha-2\gamma-1}{1+\gamma}} - \xi_g \right]
	\end{equation}
	where $\theta^g = T_{\alpha,\gamma,\tau}(G)$, 
	$
	\xi_g = \int u_{\theta^g} f_{\theta^g}^{1+\gamma} g^{1+\gamma}
	\left(\tau g^{1+\gamma} + \bar{\tau}f_{\theta^g}^{1+\gamma}\right)^{\frac{\alpha-2\gamma-1}{1+\gamma}}d\mu,
	$ 
	and 
\begin{equation}
	J_g  = \int u_{\theta^g}^Tu_{\theta^g} K'(\delta_g) f_{\theta^g}^{\alpha} g^{\gamma}d\mu  -  
	\int ((1+\alpha)u_{\theta^g}^Tu_{\theta^g} + \nabla u_{\theta^g})K(\delta_g)f_{\theta^g}^{1_\alpha}d\mu,
	\label{EQ:J_g}
\end{equation}	
with $\delta_g = \frac{g}{f_{\theta^g}}-1$ and $\nabla$ denoting the gradient with respect to $\theta$. 
\end{theorem}

\begin{corollary}\label{COR:MSDE_IF0}
Under the set-up of Theorem \ref{THM:MSDE_IF}, suppose $g=f_\theta \in \mathcal{F}$. 
Then the influence function of the  MGSDE functional (at the model) has the simpler form
\begin{eqnarray}
IF(y;T_{\alpha,\gamma,\tau}, F_{\theta}) = \left(\int u_{\theta}u_\theta^T f_{\theta}^{1+\alpha}d\mu\right)^{-1}
\left[ u_{\theta}(y) f_{\theta}^{\alpha}(y)  - \int u_{\theta}f_{\theta}^{1+\alpha}d\mu\right].
\label{EQ:S_div_IF_model}
\end{eqnarray}

\end{corollary}

\begin{figure}[h]
	\centering
	\includegraphics[width=0.45\textwidth, height=0.35\textwidth] {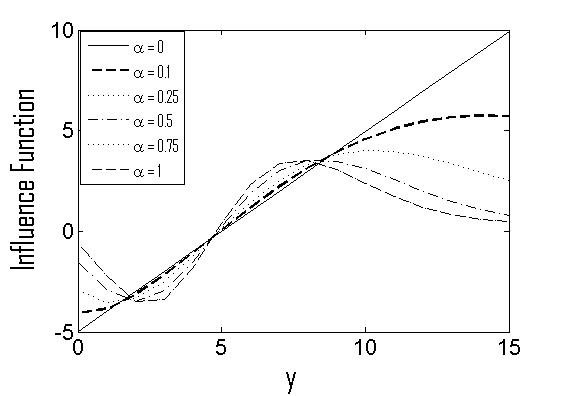}
	\includegraphics[width=0.45\textwidth, height=0.35\textwidth] {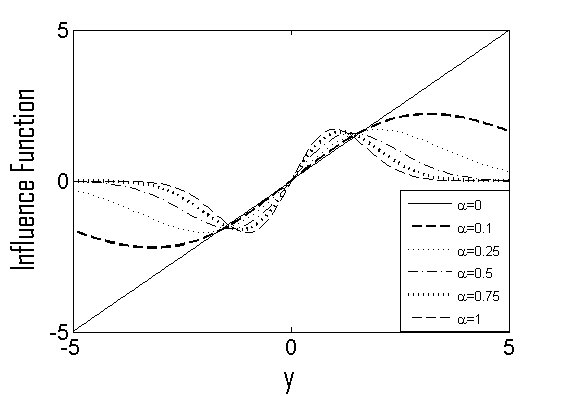}
	\caption{Influence function for the MGSDE of $\theta$ under the $Poisson(\theta)$ model at the $Poisson(5)$ 
		(first panel) and the normal $N(\theta, 1)$ model at the $N(0, 1)$ (second panel)}
	\label{FIG:IF_S_div}
\end{figure}

The most interesting and remarkable observation here is that the IF of the MGSDE at the model 
is independent of $\gamma$ and $\tau$; it depends only on $\alpha$. 
Hence the IF analysis predicts similar behavior (in terms of first order
robustness and efficiency) for all MGSDEs with same $\alpha$ irrespective of the other two parameters $\gamma$ and $\tau$. 
Also, the IFs of the MGSDEs have bounded re-descending natures except in the case $\alpha = 0$ where it is unbounded.
Figure \ref{FIG:IF_S_div} shows their natures for the Poisson-mean (discrete case) and normal-mean (continuous case).
Therefore, as per the first order IF analysis, all MGSDEs
are robust for $\alpha>0$ and non-robust at $\alpha=0$.

As suggested by one referee, the comparison of the MGSDEs, in terms of the first order IF, 
with similar existing criterion is necessarily of utmost interest here.
In this regard, we note that the IF of the MGSDE at the model is indeed the same as that of the minimum DPD estimators (MDPDEs) 
and also the MSDEs with the same value of $\alpha$ \citep{Basu/etc:1998,Ghosh/etc:2016}.
In particular, the IF of the estimators corresponding to the classical Cressie-Read PD family of the R\'{e}nyi divergence
has the form $I(\theta)^{-1}u_\theta(y)$ at the model distribution with $I(\theta)$ being the model Fisher information;
this IF is again the same as that of the MGSDE at $\alpha=0$ and 
are given by the unbounded (solid) straight lines in Figure \ref{FIG:IF_S_div}.
Therefore, in terms of the first order IF analysis, the MGSDEs with $\alpha>0$ has better robustness (bounded IF)
compared to those based on the PD of R\'{e}nyi divergence (unbounded IF)
and is similar to the MDPDEs or the MSDEs with the same $\alpha>0$.

However, in actual practice, the picture given by the first order IF analysis often 
leads to an inaccurate prediction of the actual performance of the MGSDE.
Our simulation studies in Section \ref{SEC:Illustratios} will demonstrate  
that some members of the GSD family having bounded IFs generate highly non-robust estimators;
these choices include small positive $\alpha$, $\gamma>0$ and $\tau=0$.
On the other hand, some members of the GSD family with $\alpha=0$, $\gamma<0$ and $\tau>0$ (large),
in spite of having unbounded IFs, generate highly robust estimators.
Hence, the classical first order IF analysis cannot portray the true robustness picture 
for the MGSDEs. As explained above, it may fail in both counts;
it may label strongly robust estimators as unstable, and may declare highly unstable estimators as being strongly robust.  
An appropriate second order IF analysis may provide 
a more accurate description of the distortion of the estimators due to contamination.
\cite{Lindsay:1994} and \cite{Ghosh/etc:2016} have presented some theoretical arguments, with illustrations, 
in favor of the second order influence analysis that justify robustness of estimators having unbounded first order IF.
In particular, when the first order IF is identically zero, the second order IF indicates its B-robustness,
since the linear term in the von Mises expansion of the corresponding functional vanishes. 
For some minimum divergence estimators with unbounded first order IF, like the PD family,
the second order term in the corresponding von Mises expansion becomes dominating to give more accurate 
second order bias approximations which is bounded, quantified through their second order IF;
such estimator can be thought of as second-order B-robust (from the second order bias approximation). 
\cite{Ghosh/etc:2016} have also discussed that some MSDEs, which also belongs to the larger GSD family, 
might have bounded first order IF but a dominating unbounded second order term in their von Mises expansion 
which makes them not second-order B-robust.
For brevity we do not present the second order analysis in this paper;
however this analysis reaffirms and strengthens all the illustrations and conclusions of \cite{Ghosh/etc:2016}.

\subsection{Asymptotic Properties under the Discrete Models}\label{SEC:MSDE_discrete}

Let us now describe the consistency and asymptotic distribution of the MGSDEs. 
For simplicity, we consider only the discrete models in this paper. 
Suppose $X_1, X_2,\ldots,X_n$ are $n$ independent and identically distributed 
observations from a discrete probability mass function (pmf) $g$ 
modeled by the parametric family ${\cal F} = \{ f_{\theta}: \theta \in \Theta \subseteq {\mathbb R}^p \}$ 
and let the distributions be supported, without loss of generality, on $\chi = \{ 0,1,2,\ldots \}$. 
We assume that both $g, \mathcal{F}\in\mathcal{G}$, where the dominating measure $\mu$ 
is now the counting measure over the support $\chi$.

Under this set-up, we can easily get an estimate of $g$ through the relative frequencies
defined as $r_n(x) = \displaystyle\frac{1}{n}\sum_{i=1}^n I(X_i=x)$, where $I(A)$ denote the indicator function of the event $A$.
So, we can get the MGSDE of $\theta$ by minimizing the GSD measure between two probability vectors 
$\mathbf{r}_n = (r_n(0), r_n(1), r_n(2), \ldots)^T$ and $\mathbf{f}_{\theta} = (f_\theta(0), f_\theta(1), f_\theta(2), \ldots)^T$
and hence the corresponding estimating equation is given by (\ref{EQ:S-divergence_est_equation}) 
with $\hat{g}(x)$ replaced by $r_n(x)$, the integral replaced by summation over $\chi$
and $\delta=\delta_n$ now being $\delta_n(x) = \frac{r_n(x)}{f_{\theta}(x)} - 1$.
This leads to the refined estimating equation
\begin{eqnarray}
\sum_{i=1}^n K(\delta_n(x))f_{\theta}^{1+\alpha}(x) u_{\theta}(x) &=& 0.
\label{EQ:Discrete_est_equation}
\end{eqnarray}

Now, in order to prove the asymptotic properties of the MGSDE, we consider the matrix $J_g$ as defined in (\ref{EQ:J_g})
and further define
$V_g = Var_g \left[ K'(\delta_g^g(X)) f_{\theta^g}^\alpha(X) u_{\theta^g}(X) \right], $
where $Var_g$ represents the variance under the true density $g$.
Further, we also make the following assumptions:

\begin{enumerate}
	\item[(A1)] The model family ${\mathcal F}$ is identifiable.
	\item[(A2)] The model probability mass functions $f_\theta$ have common support 
	so that the set $\chi = \{ x : f_{\theta}(x) >0 \}$ is independent of $\theta$.
	The true pmf $g$ is also supported on $\chi$.
	\item[(A3)] There exists an open subset $\omega \subset \Theta$
	for which the best fitting parameter $\theta^g$ is an interior point. Further,  the pmf 
	$f_{\theta}(x)$ admits all third order derivatives of the type $\nabla_{jkl} f_{\theta}(x)$   
	for almost all $x$ and for all $\theta \in \omega$.
	\item[(A4)] The matrix $J_g$, defined in Equation (\ref{EQ:J_g}), is positive definite.
	\item[(A5)] The quantities
	$\sum_x g^{1/2}(x) f_{\theta}^{\alpha}(x) |u_{j\theta}(x)|, ~~\sum_x g^{1/2}(x) f_{\theta}^{\alpha}(x) |u_{j\theta}(x)| |u_{k\theta}(x)|$    and\\ 
	$\sum_x g^{1/2}(x) f_{\theta}^{\alpha}(x) |u_{jk\theta}(x)|$  
	are bounded for all $j, k$ and for all $\theta \in \omega$.
	\item[(A6)] For almost all $x$, there exist functions $M_{jkl}(x)$, $M_{jk,l}(x)$, $M_{j,k,l}(x)$ 
	that dominate, in absolute value, 
	$f_{\theta}^{\alpha}(x) u_{jkl\theta}(x)$, $f_{\theta}^{\alpha}(x) u_{jk\theta}(x) u_{l\theta}(x)$ and 
	$f_{\theta}^{\alpha}(x) u_{j\theta}(x) u_{k\theta}(x) u_{l\theta}(x)$,
	respectively, for all $j, k, l$ and that are uniformly bounded in expectation with respect to $g$ and $f_{\theta}$ 
	for all $\theta \in \omega$.
	\item[(A7)] The functions $K'(\delta_g)$, $K''(\delta_g)\left(\delta_g+1\right)$ 
	with $\delta_g =\frac{g(x)}{f_{\theta}(x)} -1$ are uniformly bounded in $\theta \in \omega$.
\end{enumerate}

Then we have the following theorem stating the asymptotic properties of the MGSDE under discrete models. 
For brevity, its proof has been moved to the Online Supplementary Material.

\begin{theorem}
	\label{THM:GS-Divergence_asymptotic}
Under the set-up of discrete models as mentioned above and Assumptions (A1)--(A7), the following results hold: 
	\begin{enumerate}
		\item[(a)] There exists a consistent sequence $\theta_n$ of roots to the MGSDE estimating equation (\ref{EQ:Discrete_est_equation}).
		
		\item[(b)] Asymptotically, $n^{1/2} (\theta_n - \theta^g) \sim N_p(0_p, J_g^{-1}V_g J_g^{-1})$.
	\end{enumerate}
\end{theorem}

\begin{corollary}
Under the assumption of Theorem \ref{THM:GS-Divergence_asymptotic}, 
if $g=f_\theta \in \mathcal{F}$, then 
$n^{1/2}(\theta_n - \theta)$ has a simpler asymptotic distribution $N_p( 0, J^{-1}V J^{-1})$, 
where $J = J_\alpha(\theta) =  \int u_\theta u_\theta^T 
	f_\theta^{1+\alpha}$ and $V = V_\alpha(\theta)  = \int u_\theta u_\theta^T 
	f_\theta^{1+2\alpha} - \xi \xi^T$ with $ \xi = \xi_\alpha(\theta) 
	= \int u_\theta f_\theta^{1+\alpha}$. 
\end{corollary}

Interestingly, the asymptotic distribution of the MGSDE at the model is independent of the parameters $\gamma$ and $\tau$. 
This is also expected from their first order IFs which do not depend on $\gamma$ and $\tau$.
Hence, all MGSDEs with the same $\alpha$ (but with different $\gamma$ and $\tau$) have the same asymptotic efficiency 
at the assumed model and this efficiency is also quite high at most common parametric models for small $\alpha>0$.
This can be seen by noting the fact that the asymptotic variance and hence the efficiency of the MGSDEs at the model 
is exactly same as that of the MSDE or the MDPDEs with the same value of $\alpha$
and their high efficiencies have already been illustrated by 
\citet[][Table 1]{Ghosh:2015} and \citet[][Table 9.1]{Basu/etc:2011} respectively.

\begin{remark}[Continuous models]
For continuous models, we cannot use a simple estimate of $g$ such as the relative frequency;
the most common option is to use the kernel density estimator 
$g^*_n(x) = \frac{1}{n} \sum_{i=1}^n  W (x, X_i, h_n)$ for some kernel function $W(x, y, h_n)$ having bandwidth $h_n$.
The resulting MGSDE can then be obtained through the estimating equation (\ref{EQ:S-divergence_est_equation}) 
with $g^*_n$ in place of $\hat{g}$; but the derivations involves all the complications of kernel smoothing 
like bandwidth selection etc. 
These difficulties can be avoided through an alternative approach proposed by \cite{Basu/Lindsay:1994}
who have suggested to smooth the model density $f_\theta$ also by the same kernel as
$f^*_\theta(x) = \int W(x, y, h) dF_\theta(y)$ and minimize the GSD measure between $g^*_n$ and $f^*_\theta$.
The resulting estimating equation is given by (\ref{EQ:S-divergence_est_equation}) 
with $\hat{g}$ and $f_\theta$ replaced by $g^*_n$ and $f^*_\theta$ respectively.
See \cite{Ghosh/etc:2015} for more detail of this approach in the context of $S$-divergences
which can be extended for the GSD family in a future work. 
\end{remark}

\section{Numerical Illustrations: Performance of the MGSDEs}\label{SEC:Illustratios}

To illustrate the finite sample performances of the MGSDEs,
we generate 1000 samples, each of size $n=50$, from the $Poisson(5)$ distribution 
and compute the MGSDEs of the Poisson mean $\theta$ for different $\alpha$, $\gamma$ and $\tau$. 
The resulting empirical biases and the MSEs are reported in Table \ref{TAB:sim0} 
which show the pure-data performance of the MGSDEs.
Also, to examine their robustness, we repeat the same simulation study 
but after randomly contaminating 10\% of each sample by observations from the $Poisson(15)$ distribution; 
the corresponding empirical biases and MSEs are reported in Table \ref{TAB:sim10}.
The major findings from Tables \ref{TAB:sim0}--\ref{TAB:sim10} may be summarized as follows.

%

\begin{table}[!th]
	\centering 
	\caption{The Empirical Bias and MSE of the MGSDEs under pure data ($n=50$)}
	\resizebox{0.8\textwidth}{!}{ 	
		\begin{tabular}{r r| r r r r r|r r r r r} \hline
			& & \multicolumn{5}{c|}{Bias}& \multicolumn{5}{c}{MSE}\\\hline
			& & \multicolumn{5}{c|}{$\tau$}& \multicolumn{5}{c}{$\tau$}\\
			$\alpha$	&	$\gamma$	&	0	&	0.1		& 0.3	&	0.5	&	0.7	&	0	&	0.1		& 0.3	&	0.5	&	0.7\\\hline\hline
			0	&	0	&	$-$0.01	&	0.09	&	0.08	&	0.08	&	0.08	&	0.10	&	0.13	&	0.13	&	0.14	&	0.14	\\
			0.1	&	0	&	0.00	&	$-$0.01	&	$-$0.02	&	$-$0.04	&	$-$0.06	&	0.11	&	0.10	&	0.10	&	0.11	&	0.11	\\
			0.25	&	0	&	0.02	&	0.01	&	0.00	&	$-$0.01	&	$-$0.02	&	0.13	&	0.11	&	0.11	&	0.11	&	0.11	\\
			0.5	&	0	&	0.01	&	0.02	&	0.02	&	0.01	&	0.01	&	0.16	&	0.12	&	0.13	&	0.13	&	0.13	\\
			0	&	$-$0.3	&	$-$0.06	&	$-$0.06	&	$-$0.07	&	$-$0.09	&	$-$0.10	&	0.10	&	0.10	&	0.11	&	0.11	&	0.11	\\
			0.1	&	$-$0.3	&	$-$0.04	&	$-$0.04	&	$-$0.05	&	$-$0.05	&	$-$0.06	&	0.11	&	0.10	&	0.11	&	0.11	&	0.11	\\
			0.25	&	$-$0.3	&	$-$0.02	&	$-$0.02	&	$-$0.02	&	$-$0.02	&	$-$0.02	&	0.13	&	0.11	&	0.11	&	0.11	&	0.11	\\
			0.5	&	$-$0.3	&	$-$0.03	&	0.00	&	0.01	&	0.01	&	0.01	&	0.18	&	0.12	&	0.12	&	0.12	&	0.12	\\
			0	&	$-$0.5	&	$-$0.10	&	$-$0.10	&	$-$0.10	&	$-$0.10	&	$-$0.10	&	0.11	&	0.11	&	0.11	&	0.11	&	0.11	\\
			0.1	&	$-$0.5	&	$-$0.07	&	$-$0.07	&	$-$0.07	&	$-$0.07	&	$-$0.06	&	0.12	&	0.11	&	0.11	&	0.11	&	0.11	\\
			0.25	&	$-$0.5	&	$-$0.07	&	$-$0.04	&	$-$0.04	&	$-$0.03	&	$-$0.02	&	0.16	&	0.11	&	0.11	&	0.11	&	0.11	\\
			0.5	&	$-$0.5	&	$-$0.10	&	$-$0.01	&	0.00	&	0.00	&	0.01	&	0.23	&	0.12	&	0.12	&	0.12	&	0.12	\\
			0	&	$-$1	&	--	&	$-$0.37	&	$-$0.18	&	$-$0.10	&	$-$0.06	&	--	&	0.30	&	0.14	&	0.11	&	0.10	\\
			0.1	&	$-$1	&	0.40	&	$-$0.34	&	$-$0.13	&	$-$0.07	&	$-$0.02	&	1.83	&	0.31	&	0.15	&	0.12	&	0.11	\\
			0.25	&	$-$1	&	0.21	&	$-$0.41	&	$-$0.11	&	$-$0.03	&	0.00	&	0.65	&	0.39	&	0.19	&	0.14	&	0.13	\\
			0.5	&	$-$1	&	0.10	&	$-$0.50	&	$-$0.14	&	$-$0.02	&	0.01	&	0.35	&	0.47	&	0.24	&	0.17	&	0.15	\\
			0	&	0.5	&	0.03	&	0.02	&	$-$0.01	&	$-$0.04	&	$-$0.08	&	0.10	&	0.10	&	0.10	&	0.11	&	0.11	\\
			0.1	&	0.5	&	0.04	&	0.03	&	0.00	&	$-$0.02	&	$-$0.05	&	0.11	&	0.11	&	0.11	&	0.11	&	0.11	\\
			0.25	&	0.5	&	0.05	&	0.03	&	0.01	&	0.00	&	$-$0.02	&	0.12	&	0.11	&	0.11	&	0.12	&	0.12	\\
			0.5	&	0.5	&	0.04	&	0.06	&	0.06	&	0.06	&	0.06	&	0.14	&	0.14	&	0.15	&	0.16	&	0.16	\\
			0	&	1	&	0.06	&	0.04	&	0.00	&	$-$0.03	&	$-$0.07	&	0.11	&	0.11	&	0.11	&	0.11	&	0.12	\\
			0.1	&	1	&	0.06	&	0.05	&	0.01	&	$-$0.01	&	$-$0.04	&	0.12	&	0.11	&	0.11	&	0.11	&	0.12	\\
			0.25	&	1	&	0.07	&	0.05	&	0.02	&	0.00	&	$-$0.01	&	0.12	&	0.12	&	0.12	&	0.12	&	0.12	\\
			0.5	&	1	&	0.06	&	0.04	&	0.03	&	0.02	&	0.01	&	0.14	&	0.13	&	0.13	&	0.13	&	0.14	\\
			0	&	1.5	&	0.08	&	0.05	&	0.01	&	$-$0.02	&	$-$0.06	&	0.12	&	0.11	&	0.11	&	0.11	&	0.12	\\
			0.1	&	1.5	&	0.08	&	0.06	&	0.02	&	$-$0.01	&	$-$0.04	&	0.12	&	0.12	&	0.11	&	0.12	&	0.12	\\
			0.25	&	1.5	&	0.08	&	0.06	&	0.02	&	0.01	&	$-$0.01	&	0.13	&	0.12	&	0.12	&	0.12	&	0.13	\\
			0.5	&	1.5	&	0.07	&	0.04	&	0.02	&	0.02	&	0.01	&	0.14	&	0.13	&	0.14	&	0.14	&	0.14	\\
			1	&	1.5	&	--	&	0.03	&	0.02	&	0.03	&	0.04	&	--	&	0.17	&	0.18	&	0.18	&	0.19	\\
			\hline 
		\end{tabular}}
		\label{TAB:sim0}
	\end{table}

\begin{table}[!th]
	\centering 
	\caption{The Empirical Bias and MSE of the MGSDEs under 10\% contaminated data ($n=50$)}
	\resizebox{0.8\textwidth}{!}{ 	
		\begin{tabular}{r r| r r r r r|r r r r r} \hline
			& & \multicolumn{5}{c|}{Bias}& \multicolumn{5}{c}{MSE}\\\hline
			& & \multicolumn{5}{c|}{$\tau$}& \multicolumn{5}{c}{$\tau$}\\
			$\alpha$	&	$\gamma$	&	0	&	0.1		& 0.3	&	0.5	&	0.7	&	0	&	0.1		& 0.3	&	0.5	&	0.7\\\hline\hline
			0	&	0	&	0.96	&	0.41	&	0.29	&	0.24	&	0.21	&	1.21	&	0.42	&	0.29	&	0.24	&	0.22	\\
			0.1	&	0	&	0.58	&	0.20	&	0.13	&	0.10	&	0.06	&	0.55	&	0.22	&	0.18	&	0.17	&	0.16	\\
			0.25	&	0	&	0.31	&	0.15	&	0.11	&	0.09	&	0.07	&	0.29	&	0.18	&	0.17	&	0.16	&	0.16	\\
			0.5	&	0	&	0.13	&	0.11	&	0.09	&	0.09	&	0.08	&	0.17	&	0.16	&	0.16	&	0.16	&	0.16	\\
			0	&	$-$0.3	&	0.28	&	0.19	&	0.14	&	0.11	&	0.08	&	0.28	&	0.23	&	0.20	&	0.19	&	0.19	\\
			0.1	&	$-$0.3	&	0.20	&	0.15	&	0.12	&	0.10	&	0.09	&	0.22	&	0.19	&	0.18	&	0.18	&	0.18	\\
			0.25	&	$-$0.3	&	0.10	&	0.11	&	0.10	&	0.10	&	0.09	&	0.21	&	0.17	&	0.17	&	0.17	&	0.16	\\
			0.5	&	$-$0.3	&	0.01	&	0.08	&	0.09	&	0.09	&	0.09	&	0.24	&	0.16	&	0.16	&	0.16	&	0.16	\\
			0	&	$-$0.5	&	0.11	&	0.11	&	0.11	&	0.11	&	0.11	&	0.20	&	0.20	&	0.20	&	0.20	&	0.20	\\
			0.1	&	$-$0.5	&	0.10	&	0.09	&	0.10	&	0.11	&	0.12	&	0.21	&	0.18	&	0.18	&	0.19	&	0.19	\\
			0.25	&	$-$0.5	&	0.05	&	0.07	&	0.09	&	0.10	&	0.11	&	0.23	&	0.17	&	0.17	&	0.17	&	0.17	\\
			0.5	&	$-$0.5	&	$-$0.09	&	0.06	&	0.08	&	0.09	&	0.11	&	0.32	&	0.16	&	0.16	&	0.16	&	0.16	\\
			0	&	$-$1	&	--	&	$-$0.18	&	0.00	&	0.11	&	0.28	&	--	&	0.38	&	0.21	&	0.20	&	0.28	\\
			0.1	&	$-$1	&	0.39	&	$-$0.19	&	$-$0.01	&	0.11	&	0.25	&	1.01	&	0.41	&	0.19	&	0.19	&	0.26	\\
			0.25	&	$-$1	&	0.26	&	$-$0.31	&	$-$0.02	&	0.08	&	0.19	&	0.66	&	0.42	&	0.26	&	0.20	&	0.21	\\
			0.5	&	$-$1	&	0.11	&	$-$0.46	&	$-$0.11	&	0.08	&	0.16	&	0.30	&	0.63	&	0.26	&	0.21	&	0.23	\\
			0	&	0.5	&	2.52	&	0.29	&	0.15	&	0.09	&	0.04	&	7.02	&	0.31	&	0.19	&	0.17	&	0.16	\\
			0.1	&	0.5	&	2.22	&	0.24	&	0.13	&	0.09	&	0.05	&	5.49	&	0.25	&	0.18	&	0.16	&	0.15	\\
			0.25	&	0.5	&	1.62	&	0.18	&	0.12	&	0.08	&	0.06	&	3.03	&	0.20	&	0.17	&	0.15	&	0.15	\\
			0.5	&	0.5	&	0.55	&	0.16	&	0.13	&	0.12	&	0.11	&	0.53	&	0.20	&	0.18	&	0.18	&	0.18	\\
			0	&	1	&	3.08	&	0.29	&	0.14	&	0.08	&	0.03	&	10.34	&	0.31	&	0.19	&	0.16	&	0.15	\\
			0.1	&	1	&	2.95	&	0.24	&	0.13	&	0.08	&	0.04	&	9.52	&	0.26	&	0.18	&	0.15	&	0.15	\\
			0.25	&	1	&	2.70	&	0.19	&	0.11	&	0.08	&	0.05	&	8.02	&	0.21	&	0.17	&	0.15	&	0.15	\\
			0.5	&	1	&	2.05	&	0.13	&	0.10	&	0.08	&	0.07	&	4.80	&	0.19	&	0.16	&	0.16	&	0.15	\\
			0	&	1.5	&	3.29	&	0.27	&	0.14	&	0.08	&	0.03	&	11.71	&	0.30	&	0.18	&	0.16	&	0.15	\\
			0.1	&	1.5	&	3.23	&	0.23	&	0.12	&	0.08	&	0.04	&	11.27	&	0.25	&	0.17	&	0.15	&	0.15	\\
			0.25	&	1.5	&	3.11	&	0.18	&	0.11	&	0.08	&	0.05	&	10.50	&	0.22	&	0.17	&	0.15	&	0.15	\\
			0.5	&	1.5	&	2.81	&	0.13	&	0.09	&	0.08	&	0.07	&	8.69	&	0.19	&	0.17	&	0.16	&	0.16	\\
			1	&	1	&	--	&	0.10	&	0.10	&	0.10	&	0.11	&	--	&	0.19	&	0.19	&	0.19	&	0.19	\\
			\hline 
		\end{tabular}}
		\label{TAB:sim10}
	\end{table}

\begin{enumerate}
	\item Under pure data, the absolute bias is minimum at the MLE ($\alpha=\gamma=\tau=0$) as expected. 
	However most other members of the GSD family generate competitive results.
	
	\item The MSE under pure data is also minimum at the MLE and increases with $\alpha$ for any fixed $\lambda$ and $\tau$;
	this is  expected from their (theoretical) asymptotic variance at the model. 
	However, many MGSDEs again have quite competitive MSE values and hence the loss in efficiency 
	under pure data is not a very serious concern for them.
	
	\item Under the contaminated scenario, the bias and MSE become quite high for the MLE with respect to the pure data case. 
	But several MGSDEs are stable and do not depart significantly away from their values in pure data. 
	
	\item The stable MGSDEs generally correspond to the larger choices of $\alpha$ and $\tau$.
	In particular, both the bias and MSE generally show a decreasing pattern with increasing $\alpha$ or $\tau$.
	
	\item 
	For larger values of $\tau$ and $\alpha$, there is no significant effect of $\gamma$ on the robustness of the MGSDEs.
	But for small $\alpha$ or $\tau$, 
the MGSDEs with $\gamma<0$ generate stable bias and MSEs under contamination indicating  a strong degree of robustness, 
whereas those with $\gamma>0$ become even more unstable than the MLE.
\end{enumerate}

Therefore, many of the newly developed MGSDEs are highly robust under data contaminations 
and yield only a small loss in efficiency under pure data.
Further, the best MGSDEs in terms of both bias and MSE correspond to the tuning parameters  
$0.1\leq\alpha\leq 0.25$, $1\leq\gamma\leq 1.5$ and $0.5\leq\tau\leq 0.7$;
roughly they appear to provide the best compromise between efficiency at the model and robustness under data contamination.  
Interestingly, these GSD measures do not belong to any of the existing divergence families 
like PD, GKL or SD; in fact they are far separated from the existing ones. 
Hence the development of this larger GSD family does not limited only to a theoretical generalization in an academic interest. 
Rather, they produce new MGSDEs yielding more robust inference in real practice with contaminated data,
compared to the other existing minimum divergence estimators.


\section{Concluding Remarks}\label{SEC:Conclusion}

In this paper, we have discussed the divergence based MATs and their link with some new divergence families.
We have demonstrated the development of the SD family and a new larger superfamily (GSD) of divergences from suitable MATs. 
We have also discussed several interesting properties of this new GSD family 
and its potential application in robust parametric inference.
In this pursuit, we have demonstrated the limitations of the first order IF in assessing their robustness.

However, the GSD family have some identical members and their  topological characterization will be an interesting future research.
Its application to different inference problems will also provide a great value addition in robust statistics. 
Further studies may generalize the studied connection between the MAT and resulting divergence family. 
Up to this point, each time the new divergence has been generated from a MAT, it  appears to have provided better 
trade-off between good efficiency and robustness properties compared to the existing ones. 
So, a relevant question is how long we can extend this process to generate even larger superfamilies of divergences.
We hope to pursue some of these extensions in our future endeavors.

%

\bigskip\noindent
\textbf{Acknowledgments:} 
The authors gratefully thank the Editor, an Associate Editor and two anonymous referees for their useful comments
which led to an improved version of the manuscript.

\end{document}